# On Design-time Security in IEC 61499 Systems: Conceptualisation, Implementation, and Feasibility

Awais Tanveer, Roopak Sinha, Stephen G. MacDonell
*School of Engineering, Computer and Mathematical Sciences,*
*Auckland University of Technology, Auckland, 1010*
{awais.tanveer; roopak.sinha; stephen.macdonell}@otago.ac.nz

**Abstract**

*Cyber-attacks on Industrial Automation and Control Systems (IACS) are rising in numbers and sophistication. Embedded controller devices such as Programmable Logic Controllers (PLCs), which are central to controlling physical processes, must be secured against attacks on confidentiality, integrity and availability. The focus of this paper is to add design- level support for security in IACS applications, especially around inter-PLC communications. We propose an end-to-end solution to develop IACS applications with inherent, and parametric support for security. Built using the IEC 61499 Function Blocks standard, this solution allows us to annotate certain communications as 'secure' during design time. When the application is compiled, these annotations are transformed into a security layer that implements encrypted communication between PLCs.*

*In this paper, we implement a part of this security layer focussed on confidentiality, called Confidentiality Layer for Function Blocks (CL4FB), which provides a range of encryption/decryption and secure key exchange functionalities. We study the impact of using CL4FB in IACS applications with real-time constraints. Through a case study focussing on protection functions in smart-grids, we show that varying levels of confidentiality can be achieved while also meeting hard real-time deadlines.*

## 1. INTRODUCTION

Industrial Automation and Control Systems (IACS) have significantly changed the way we operate manufacturing plants, transportation systems, facility management and monitoring systems [1]. Of the many requirements relating to system quality, security is of chief importance in IACS. Legacy IACS were often confined within factory walls which is no longer the case. Now, they actively communicate with the outside world, leaving them more vulnerable to the security threats than ever before.

Although the top layers of most IACS are secured using network security mechanisms, devices at lower levels such as Programmable Logic Controllers (PLC), Remote Terminal Units (RTU), sensors and actuators are often left susceptible to attacks. Such small embedded devices control and process inputs/outputs from sensors and actuators. As they tend to have constrained resources and are designed to provide simple and reliable operations, cryptographic requirements for security purposes, like encryption and authentication mechanisms, are often overlooked which makes them highly vulnerable to attacks [2]. Stuxnet [3], which specifically attacked PLCs in nuclear infrastructures, is a major example showing the extent of the damage possible if such devices are improperly secured. A vulnerability report on industrial control systems [4] shows that more than 90% of hosts connected to IACS have vulnerabilities ready to be exploited.

The IEC 61499 [5] is a standard for designing and developing IACS applications. It is an event-driven architecture that uses Function Blocks (FB) connected to each other to form a chain of events. An FB provides an interface consisting of I/O events and data ports. There is a steady flow of industry practitioners adopting IEC 61499 to develop distributed ap- plications. Research on security considerations in IEC 61499 based distributed applications is in the very early stage. There is also a scarcity of solutions that deal with full-scale security implementation regarding confidentiality, integrity and authentication in IEC 61499 distributed applications. The solitary solution that we were able to find is the proposed authentication layer [6] which is more conceptual and is yet to be evaluated. The solution discusses the importance of secure key exchange but falls short of providing the enabling implementation. The contemporary research contains no solution providing confidentiality and integrity services to IEC 61499 based applications, therefore exposing communication between control devices make them susceptible to attackers. Also, existing solutions to secure communication like OPC Unified Architecture or Secure Socket Layer demand more processing power in small embedded devices which is not always available.

In this research, we propose a method where IEC 61499 distributed FB data links can be annotated with security requirements at the design-time. Then, pre-configured security mechanisms, forming a security layer, are added automatically at compile time to ensure secure communications between distributed slices of the application. In this paper, we have also implemented a security layer focussed only on confidentiality. In the proposed layer, called *Confidentiality Layer for Function Blocks* (CL4FB), we implement encrypted data communications using Advanced Encryption Standard (AES) [7]. We also implement the Diffie-Hellman Key Exchange (KE) algorithm for setting up and renewing secure communication sessions. These features are encoded using IEC 61499 compatible composite and service-interface function blocks. We have conducted a study to explore the feasibility of the proposed CL4FB, and its impact on system functions with strict timing constraints. Through a case study of protection functions in smart-grids, we find that while the CL4FB does introduce latency, most real-time constraints can be met while also ensuring some level of confidentiality. The key contributions of this paper are: 1) *Secure links*, a design-time mechanism that allows designers to annotate secure communications. 2) A *confidentiality based security layer* (CL4FB) and its implementation to support secure links. 3) A *feasibility study* showing how the security-performance trade-off can be managed at design-time.

The rest of this paper is organised as follows: Section 2 presents background and a review of related literature. Section 3 discusses the notion of secure data links for an IEC 61499 application and subsequently proposes the CL4FB model and implementation. Section 4 implements our solution on an IACS application for protecting smart-grid power distribution networks. Section 5 presents the feasibility regarding latency that is incurred by adding CL4FB to the case study. Finally, section 6 provides the conclusions and future directions.

## 2. BACKGROUND AND LITERATURE REVIEW

*A. Industrial automation and control systems*
IACS are highly distributed and heterogeneous systems. They are an integral part of safety-critical cyber infrastructures around the world [8]. These systems contain technologically diverse sensors, actuators, controllers, communication channels and monitoring systems which leads to high complexity of design and development. Controller devices play the role of an arbiter between other components by collecting data from sensors, processing it and subsequently driving the actuators. They also interface with Supervisory Control And Data Acquisition (SCADA) Systems for monitoring purposes [9]. Controller devices include PLCs and RTUs that are embedded devices often with limited processing power [9].

*B. IACS and Security*
Increasing emphasis on automating industrial processes means that the security of such infrastructures is of utmost importance. Several attacks (German nuclear plant, Dragonfly and Stuxnet) on IACS with devastating consequences have been reported in the last decade [10]. Researchers and practitioners are facing a race against time to protect and secure critical infrastructure, as the automation technology for industrial processes is taking giant strides in its scale and reach.

A 2016 Industrial Control Systems (ICS) vulnerability report by Kaspersky Labs [4] shows a grim picture regarding ICS security. Their main findings show that the number of exploitable vulnerabilities in ICS components are growing with time. Most of the vulnerabilities (49%) found are of a high-risk level, and 5% of the vulnerabilities found in 2015 were yet to be patched at the time of publishing of this report. ICS components that are available through the internet are using insecure protocols such as Http, Telnet, SNMP, etc. Most of these components can be accessed externally and combining this phenomenon with the use of insecure protocols, has resulted in 92% of hosts deemed to be insecure and vulnerable. The report also finds out that a wide range of industries is affected by the lack of security measures taken against exploitable vulnerabilities.

Citing [4], [11] raises concerns about the state of hardware security and the need for appropriate measures to be taken by vendors. These works highlight the role of cryptography in securing hardware devices, but the ever-changing landscape of ubiquitous computing demands a more extensive approach. It includes robust architectures and frameworks with security requirements integrated by-design. Lack of resources (e.g. memory and processing power) in such devices also means that they have to bank on lightweight security solutions which otherwise, may lead to attacks such denial of service where an attacker overwhelms its target. Therefore, robust measures should be taken when considering securing such controller devices. A system Protection Profile (PP) developed by the Process Control Security Requirements Forum (PCSRF) [9] specifies the minimum security requirements for field devices such as PLCs and RTUs for SCADA systems in medium robustness environments. The PP also requires that if the Target Of Evaluation (TOE) (Field device) implements cryptographic functionality, then it must comply with FIPS 140-2 [12] requirements at certain levels. One of the cryptographic requirement is the provision of encryption services that are also being advocated in the British Standard for Industrial Communication Networks — Network and system security (IEC-62443-3-3:2013) [13].

*C. Security in IEC 61499 based applications*
The IEC 61499 standard provides mechanisms to develop applications for industrial distributed systems. It specifies a distribution model for deploying an application on multiple devices and a management model for managing resources within a device for compliant applications. It is an event-driven architecture that uses interconnected function blocks (FBs) to form chains of events. FB have interfaces containing I/O events and data ports [5]. It relies on an Execution Control Chart (ECC) to implement internal logic through associated algorithms. IEC 61499 allows distributing the functionality of a software module across multiple controller devices. Moreover, it helps in the development of the distributed application by focussing on re-usability and reconfigurability through design- time abstractions of underlying hardware.

The distribution model of IEC 61499 specifies that a sub-application type containing multiple FBs can be deployed on one or more devices. The distribution is seamless as far as the architecture is concerned, but the distributed FBs communicate using a communication network. The standard specifies interfaces through communication FBs (that are a type of SIFBs), but the underlying properties of a communication channel are not in the scope of the standard. Securing such channels is very desirable as IACS are susceptible to a variety of cyber attacks [2]. In a distributed IEC 61499 application, FBs communicating over a communication network regularly need to send I/O parameters. An attacker can disrupt the services by taking advantage of communication layer vulnerabilities.

A proposal to achieve integrity and authenticity in the context of IEC 61499 using the Universal Message Authentication Code lightweight algorithm for message authentication through an Authentication Layer has been described in [6]. It utilises the publisher/subscriber model of IEC 61499 in multi-cast mode, i.e. a publisher can communicate with multiple subscribers. It assumes that the keys have already been trans- mitted securely before the start of the session. It proposes the use of FORTE [15] IEC 61499 runtime for the implementation purposes. A similar approach to protect message integrity between PLCs hosting IEC-61131-3 [16] — a predecessor of IEC 61499 — based applications is presented in [17].

Although message integrity and authenticity are priorities in IACS environment, lack of confidentiality in distributed controller communication can also disrupt the services provided by these devices. The confidentiality goal is achieved by encrypting the data. Encryption ensures that an unauthorised party does not get to read the private information. In the absence of the encryption, an attacker can eavesdrop on the data to alter it in ways that can lead to activities such as unauthorised monitoring and traffic analysis. It is also true in case of IEC 61499 based applications. For example, unencrypted data exchange between two FBs deployed on multiple devices in a sensitive environment can cause a breach of private data leading to the revelation of critical parameters to the attacker. Although, an attempt to introduce confidentiality in PLCs has been made in [14] but it lacks the support for distributed applications. In this research, we will largely focus on providing cryptographic services to a distributed IEC 61499 based application that will help in achieving the goal of confidentiality.

## 3. SECURE SIFBS FOR IEC 61499 APPLICATIONS

The IEC 61499 standard specifies different types of FBs. All FBs have unique interfaces consisting of I/O events and data ports. The execution of a Basic FB (BFB) relies on an ECC which is a finite state machine that implements internal control logic using associated algorithms. Composite Function Blocks (CFB) contain networks of FBs to achieve more complex functionalities. A Service Interface Function Block (SIFB) is like a device driver that provides services to IEC 61499 applications, such as network communication while hiding the underlying complexities.

Fig. 1 shows how parts of an IEC 61499 application can be deployed over separate PLCs, each hosted within a supported runtime. The communication SIFBs may be inserted where needed to ensure event and data flow between these runtimes. However, data transmitted by such SIFBs over a network is susceptible to attacks unless the transmission medium is secure. Heavy-weight solutions such as IPSEC are often not feasible due to the limited processing power of the PLCs [18]. Also, leaving the selection of the communication medium until deployment time is not ideal, especially as the configuration of the system over PLCs can change. Currently, secure communications are treated no differently to any other communication during design time.

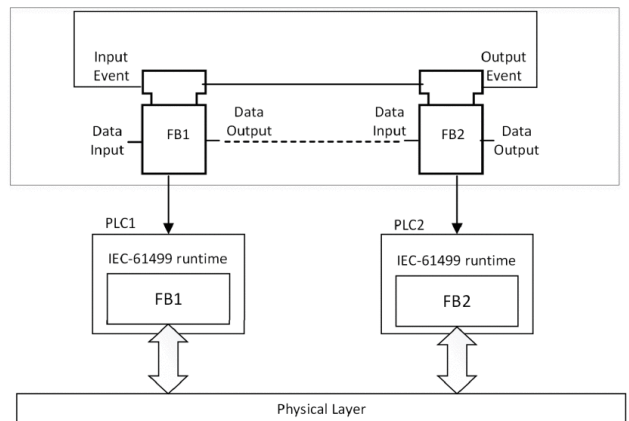

Fig. 1. IEC 61499 application FB distribution concept

### A. Secure links for distributed IEC 61499 based applications

We propose a framework that allows designers to identify secure data links within an IEC 61499 application during the design time. At the compile time, such links are automatically translated into pre-configured application logic and supporting SIFBs, to make the links secure. Formally, a secure link `sl` is defined as:

$$sl = (d\_con, sec\_goal, alg, params)$$

where `d_con ∈ D_Conns` is any link between the data ports of an FB application [5], `sec_goal ∈ {Confidentiality, Integrity, Availability}` identifies the type of security required, `alg ∈ Algs(sec_goal)` is an avail- able algorithm for the type of security required, and params is a set of `parameters` required for configuring algorithm alg. For instance, a data link requiring confidentiality can be annotated with a specific confidentiality algorithm and related parameters such as key size, encryption mode etc. Similarly, if integrity is a required goal for a link, it can be annotated by specifying algorithms such as HMAC and corresponding parameters. Secure data links essentially translate to annotations on specific data-links in a FB application. These annotations, namely `sec_goal, alg,` and `params` are used by the compiler to automatically include and configure the required security code from an existing FB library. This framework provides improved usability and abstraction by preventing the need to include and configure security blocks explicitly at design time. Fig. 2 illustrates an example of secure IEC 61499 FB data link. `@secure` is the keyword to mark the secure data link. The first parameter `C` stands for Confidentiality that is a security goal. The second parameter `AES` is the name of the encryption algorithm. The last parameter may accommodate varying numbers of arguments required for the algorithm used in the second parameter. The event connection/link (CNF to REQ) does not carry the data but only acts as a trigger. Therefore, encrypting event links is not applicable especially in this case.

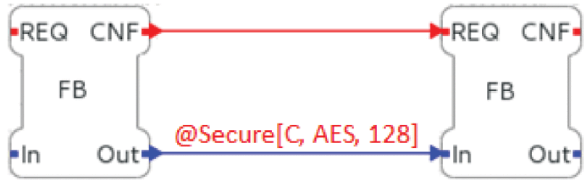

Fig. 2. A secure FB data link with annotation

For the compiler to generate secure links, a set of enabling FBs and SIFBs must be available, in the form of a library.

In the following sections, we demonstrate an FB library for proving security, focussed on confidentiality. The library can be extended in the future to provide other security elements such as integrity, availability, authenticity etc.

### B. Confidentiality Layer for Function Blocks (CL4FB)

All secure data links in an application are compiled into a logic that supports secure communications. The logic, implemented using IEC-61499 FBs, ultimately forms an independent layer whose chief aim is to secure data communications between distributed parts of the application. In this paper, we propose a confidentiality-based security layer built on the IEC 61499 distribution model. The library, called Confidentiality Layer for Function Blocks (CL4FB), provides the services related to confidentiality. It can be envisioned as a secure tunnel through which data from one FB is transmitted to another FB over a non-secure medium.

An alternative way to ensure security is to secure the medium over which PLCs communicate. It can be complex, especially in the case of confined but critical industrial networks. It may include deploying security protocols, e.g., a Transport Layer Security (TLS) or IP Security (IPsec) if Ethernet-based communication is required, the set up of which may not be feasible in the PLC network topology. Constraints like power consumption and timing can limit the usage of this alternative approach. By incorporating security within IEC 61499 distributed applications, we can avoid the overhead of using a secure medium, and also make the application more self-contained and portable.

CL4FB consists of encryption and decryption logic implemented using FBs. Fig. 3 shows the abstract configuration of FBs in a distributed application forming CL4FB. The choice of implementation is the prerogative of the designer and the security level they desire.

In this research, we assume that a block cypher is more appropriate in the current context as the amount of data to be processed is known in most cases. In the case of stream cyphers, the amount of data to be encrypted is usually unknown. When choosing a block cypher, the primary criteria to consider is the trade-off between security and performance. We choose to adopt the well-known and widely accepted AES algorithm for secure data communications.

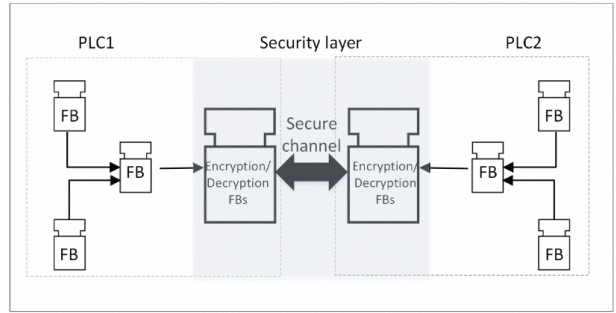

Fig. 3. Proposed Security layer in an IEC 61499 distributed application

Key expansion is an integral part of block cyphers. It is the foremost step performed to expand a smaller key into a larger key that is used in a subsequent Feistel network [19] that is common to block cyphers. It is a one time-operation that is performed before an encryption session until the key is replaced. In the proposed CL4FB, the key expansion process can be realised as a separate FB because incorporating key expansion in each cycle of encryption induces unwanted latency in the encryption/decryption process. A designer must make sure that the key expansion is performed before the actual encryption process starts on all controller devices where the FBs of an IEC 61499 application are distributed.

### C. (SI)FBs for the CL4FB

The most common method to transfer data between distributed FBs is through publisher/subscriber SIFBs [5]. Two data transmitting FBs need a pair of publisher-subscriber SIFBs to send data over the network. A publisher is used for an FB providing encryption services to send cypher text over the network. Similarly, decryption FB uses a subscriber SIFB. Publisher/Subscriber pair may be conceived as an implicit part of the CL4FB providing confidentiality services. A more detailed view of CL4FB can be seen in Fig. 4. In the case when a distributed application needs to send secure data in a bidirectional manner, another set of publisher/subscriber SIFBs along with encryption/decryption FBs can be used in a flipped configuration. An alternate way of aiding a simpler design for an IEC 61499 distributed application is to implement both encryption (publish/send) and decryption (subscribe/receive) algorithms in a single FB. It can be efficiently realised by adding a predicate variable to the FB interface and subsequently incorporating relevant logic in the ECC of the FB.

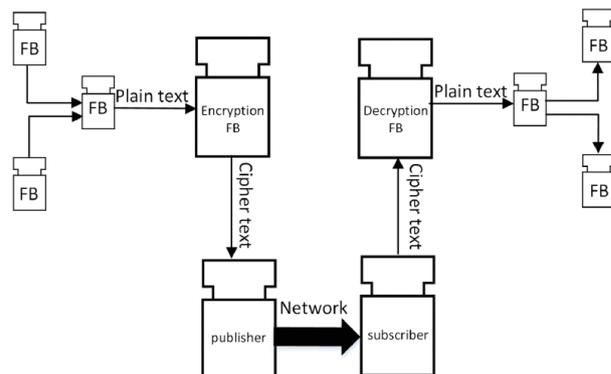

Fig. 4. An internal view of IEC 61499 distributed application with CL4FB

Fig. 5 shows the event and data I/O interfaces of the FBs proposed for encryption services and designed for the I/O requirements of the block cyphers. IEC 61131-3 data types are used to define I/O data. The service is started when the `REQ` event is triggered, and triggering of a `CNF` event indicates that the services have been completed. In the encryption FB, *pt* is a 16-byte array which makes a block of plain text. *ksize* denotes the size of key to be used in a block cypher. The most common lengths are 128, 192 and 256 bits. It is an important variable because the length of the key determines the number of rounds to be executed in a Feistel network block cypher. *expkey* is the expanded key that is the outcome of the key expansion process. Its size depends upon *ksize*. Finally, *ct* is the cypher text which is the output of the encryption process. All the data I/O parameters described here remain the same for decryption FB except that it takes cypher text (*ct*) as an input and decrypts it to produce the plain text (*pt*). The key expansion FB simply takes key and key size as input parameters and outputs an expanded key which is then fed into the encryption and decryption FBs.

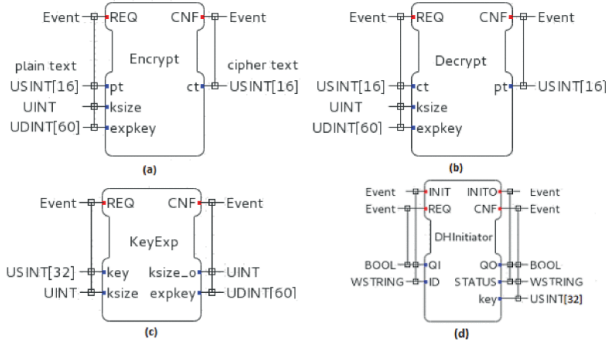

Fig. 5. CL4FB event and data interfaces: (a) FB for encryption services (b) FB for decryption services (c) FB for Key expansion process (d) Proposed Secure KE SIFB

*D. Secure Key Exchange*

The security of an encryption algorithm depends upon its keys. An attacker can get hold of the key if the medium used to transfer the key is not secure. There has been a significant amount of research regarding methods of secure Key Exchange (KE) and distribution [20]. Participating controller devices must have the same key for the encryption and decryption process if using symmetric key block cyphers to secure the communication between distributed FBs. Therefore, we also propose a supplementary secure KE process implemented in the form of an SIFB that may be slotted in as a precursor to the actual encryption process.

Cryptographic KE protocols such as Diffie-Hellman KE [21] require a Random Number Generator (RNG) on each participant device, implemented as a Pseudo-Random Number Generator or a True Random Number Generator. In either case, an FB implementing KE protocols requires accessing the entropy pool for RNG. Moreover, Diffie—Hellman KE also requires the transfer of public parameters over the network. Therefore, an SIFB is used to implement the KE protocol as it can provide underlying system services for RNG and network operations.

Fig. 5 (d) presents proposed SIFB for secure KE in an IEC 61499 distributed application. The SIFB can act both as an initiator and a responder. Events in KE SIFB are standard `INIT,INITO,REQ` and `CNF.INIT` can be used to initialise KE protocol's context structures and generate random numbers from the entropy pool. Data input *QI* is used to determine the role of the particular SIFB, i.e. initiator or responder. Input *ID* is a standard publisher/subscriber parameter that is used to assign a unique pair of IP and port number. *STATUS* shows the current status of the service at each phase of KE process. A public value computed by each SIFB is transferred to the participant peer SIFB using system's underlying network services. The public value is used to compute the symmetric shared secret key in for Diffie—Hellman KE. Finally, *Key* output parameter is the generated symmetric key that can subsequently be used by encryption/decryption FBs. A com- prehensive data flow model involving (SI)FBs of CL4FB along with secure KE can be seen in Fig. 6.

The proposed security (SI)FBs maintain secure sessions along with security policies in an IEC 61499 environment. The encryption/decryption FBs and the KE SIFB offers flexible interfaces to implement a variety of block cyphers and KE protocols. Sessions may be maintained by putting a time constraint on the usage of a particular key after which, re- keying and secure KE must be performed, and a new key is distributed amongst participating devices. Moreover, multiple block cyphers may be implemented in the proposed encryption/decryption FB encapsulations that opens up the possibility of dynamic set of options to be used in each session. However, such mechanism is out of scope for this paper.

## 4. CASE STUDY: SECURE SMART-GRID PROTECTION

In this section, we demonstrate the applicability and viability of CL4FB using an IEC 61499 based solution for protection and control functions in electric power distribution [22]. Here, IEC 61499 is used for three protection functions, earth fault, over-current protection and differential protection. A similar concept of Intelligent Fault Management is also discussed in [23] where trip signals from protection functions are sent to circuit breakers using User Datagram Protocol connection.

The over-current protection function safeguards the system against issues like short circuits or overloaded lines. It trips the circuit breaker when the currents surmount a threshold value of 100A. It must trip the circuit within 600ms as specified in the referenced research. Similarly, a differential protection function protects the power transformer from internal malfunction when the difference between two currents in the transformer exceeds 1A, triggered within 5ms of the fault occurrence. We use these timing requirements as benchmarks to show the feasibility of CL4FB in this research.

Electric power distribution requires high security due to its widespread utilisation and pivotal role in critical infrastructures. The possibility and nature of cyber and physical attacks on electric power delivery system have been discussed in [24].

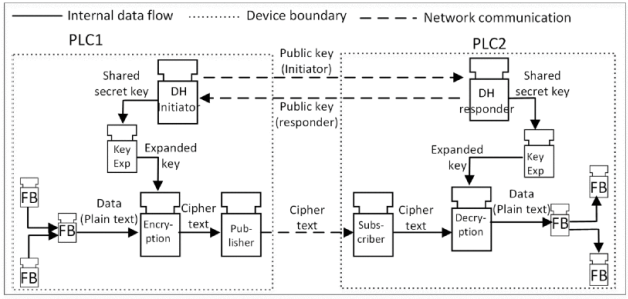

Fig. 6. An all-inclusive model of CL4FB with key exchange

Fig. 7 shows IEC 61499 implementation scenario of three protection functions described earlier, designed to be implemented in designated Intelligent Electronic Devices (IEDs). The IED sends a trip signal to a circuit breaker when an anomaly occurs in current or voltage levels. Circuit break is perceived to be a separate IED as well. All the devices on the left side of the figure communicate to circuit breaker using the Ethernet. Each left-to-right link is annotated with security parameters assessed to fulfil the timing requirements. That is, AES128 with fewer rounds is appropriate for strict timing requirement of 5ms for differential protection function while AES192 or AES256 is well-suited for the slightly relaxed timing requirement of overcurrent protection. As far as the annotations on the links are concerned, these are envisioned to be part of an automated security framework for IEC 61499 based IDE and run-times where annotations may help compilers to generate and configure FBs according to the provided security parameters. In this research, we have executed the same process manually.

Fig. 8 demonstrates a CL4FB implementation to secure the communication between protection FBs and the circuit breaker. The dotted outline represents the domain of the layer. It consists of two CFBs named `CLSender` and `CLRecv` responsible for sending and receiving the encrypted trip signal respectively. FBs deployed in an IED are shown in corresponding colours. In this case, three `CFSender` FBs—deployed on designated IEDs—are required to carry the encrypted trip signal over the network from respective protection function and deliver it to the circuit breaker. *ID* in `CLSender` and CLRecv input interfaces is the combination of multicast IP host group and the port. Due to the timing requirements of protection functions described earlier, `CLSender` and `CLSender_1` encrypt the trip signal with a 128 bit key (via *keysize* interface), so they share the same IP:Port combination. The `CLRecv` CFB receives the trip signal and decrypts it before sending it to the circuit breaker. Similarly, `CLSender_0` and `CLRecv_0` communicate at different IP:Port combination using a 256- bit key but they can also use 128 or 196-bit keys due to the slightly relaxed requirements of overcurrent protection. The *rekey* parameter specifies the time interval after which a new key must be negotiated between communicating IEDs.

Fig. 9 represents internal FB network of `CLSender` CFB. `DHInitiator` SIFB and `AESKeyExp` FB are executed once per session; the length of which is determined by *rekey* input parameter associated with `E_CYCLE` FB. Expanded key from `AESKeyExp` is fed into

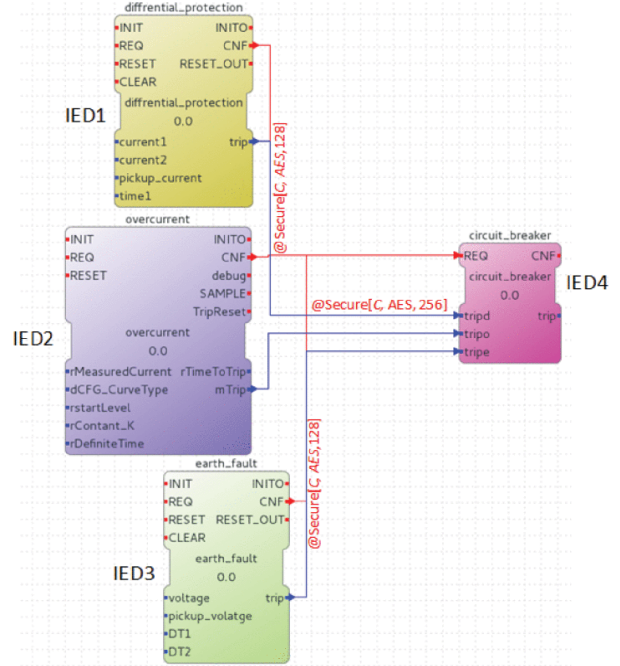

Fig. 7. An IEC 61499 system for implementing protection functions

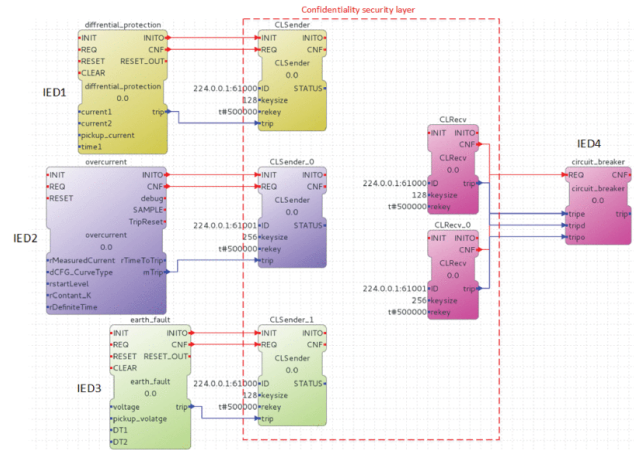

Fig. 8. Implementation of CL4FB for protection functions

`AESEncrypt` while `ConvertToArray` is another FB that converts the boolean value of the trip signal into an input plaintext block of 16 bytes required by AES. Resulting cyphertext is sent to the circuit breaker over the network using a publisher SIFB. Similarly, Fig. 10 serves an internal FB network of `CLRecv` CFB that is deployed in circuit breaker IED. Cyphertext sent over the network is received by the subscriber SIFB and forwarded to `AESDecrypt` FB for the decryption. The resulting plaintext block is converted back to a boolean trip value.

Establishment of a secure channel in this case study is only the demonstration of the capability of the proposed CL4FB. Having said that, using larger aliases for boolean trip values may enhance the security. In the next section, we test the feasibility of our approach by deploying and executing the FBs in an IEC 61499 runtime environment.

## 5. FEASIBILITY ANALYSIS

The trade-off between a system's performance, usability and functionality at the expense of security has been a topic of interest in the research community [25]. The trade-off between the performance and security is well known, due to their competing needs for the processing power. In this section, we evaluate how the proposed security library affects the performance, characterised by latency, for the protection system as discussed in section 4.

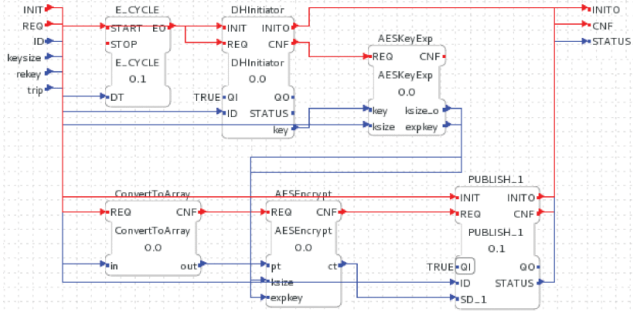

Fig. 9. CLSender CFB internal FB network

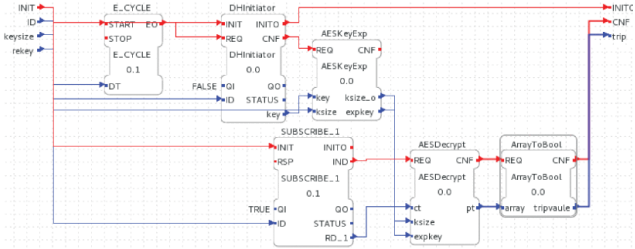

Fig. 10. CLRecv CFB internal FB network

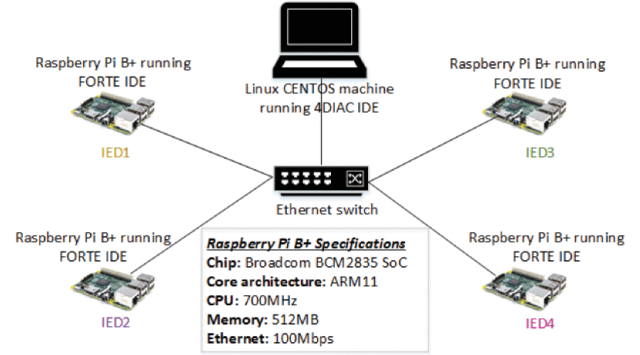

Fig. 11. Experimental setup

To realise the CL4FB (SI)FBs, we have used the 4DIAC Integrated Development Environment (IDE) (version 1.8.4) that provides a platform to develop distributed applications according to IEC 61499 specifications. It also provides FORTE which is an open-source portable Runtime Environment (RTE) for executing IEC 61499 applications. FORTE (version 1.9.0.M1) was used to deploy and execute FBs on Raspberry Pi B+ boards to simulate IEDs. Fig. 11 illustrates the experimental setup. A Linux machine (CENTOS 7.2) runs the 4DIAC IDE that is used to design and deploy IEC 61499 distributed applications. The deployment process consists of sending FB constructs including the information related to distributed operations, to FORTE RTE instances each running on a separate Raspberry Pi. The distributed FBs deployed on the FORTE RTE communicate with each other via Ethernet. A Network Time Protocol (NTP) server was established in a LAN environment to ensure the synchronisation of the systems clocks on each Raspberry Pi. A less than 1 ms NTP offset was achieved due to little congestion in the LAN implemented only for the NTP purposes. We deem NTP server and client configurations to be out of the scope this paper. For latency measurement, an SIFB `TimeStampRecorder` recording UNIX timestamps in milliseconds was developed. A time stamp $t_1$ was obtained by placing `TimeStampRecorder` before the `AESEncrypt` FB in each of the `CLSender` CFB. Similarly, time stamp $t_2$ was obtained by putting `TimeStampRecorder` right after the `AESDecrypt` FB in each of the `CLRecv`. The time stamp $t_1$ was sent over the network to the Raspberry Pis having `CLRecv` CFBs using a separate publisher/subscriber pair. The latency (L) is then calculated by $L = t_2 - t_1$. Using this setup, we measured the latency of CL4FB to analyse their usability in IACS systems. AES block cipher is implemented in encryption/decryption FBs using Electronic Codebook Mode (ECB). In the proposed SIFB for secure KE, Diffie—Hellman KE method is implemented along with necessary network communication using multicast mode. A stub application was created to generate the plain text which is then fed into the encryption FB.

TABLE I. LATENCY OF FBS FOR THE PROPOSED CL4FB

| Latency without encryption process | 1-2 ms | | |
|---|---|---|---|
| Configuration | AES128 | AES192 | AES256 |
| Single controller device | 2-3 ms | 3 ms | 3-5 ms |
| FBs distributed on multiple IEDs | 5-6 ms | 8-9 ms | 10 ms |

Calculating delays is an important factor to consider when design distributed applications where response time to an event is critical [26]. Table I shows the latency induced by the CL4FB. The results are obtained by noting the minimum- maximum latency values over 100 cycles for each scenario. The first row of the table shows the processing time taken by encryption and decryption FBs, i.e. time taken from when encryption FB's `REQ` event is triggered to the triggering of a `CNF` event of decryption FB that outputs the decrypted plain text. Both FBs are deployed on the same machine in this instance. However, the second set of results present the latency of the encryption process when FBs are distributed on multiple devices. Resulting values are slightly higher due to network communication. The processing time for key expansion is not included as it is a one-time process that is not required for the encryption of each block.

The results in Table I indicate that AES with 128, 192 and 256-bit key may well be suitable for overcurrent protection function because of its lenient timing requirements. On the other hand, only AES128 may satisfy the requirement of differential and earth fault protection functions in the current experimental setup. It can be seen that a 5 ms threshold value is barely achieved even by AES128 that may cause problems and damage the equipment by not tripping the circuit in the required amount of time. However, execution platforms more powerful than the deployed Raspberry Pi B+ may produce better results to satisfy the timing requirements of differential and earth fault protection functions. Therefore, a designer can consider these results as a clue to assess the feasibility of introducing CL4FB into IEC 61499 distributed applications. Subsequently, it will also help in the selection of appropriate cryptographic transform to minimise the trade-off between security and latency. For example, a light-weight encryption algorithm can greatly help to reduce the latency when security is a desired feature. Also, we believe that selecting an IEC 61499 dedicated IED can eliminate the overhead caused by the general-purpose operating system in a Raspberry Pi. The timing constraints noted in [27] regarding smart-grid based Future Renewable Electric Energy Delivery and Management (FREEDM) systems project, lie within the range of results.

Similar to the key expansion operation, secure KE is involved whenever the key needs to be updated, after a specific interval of time depending on the security policy. Latency induced by secure KE depends upon external factors such as generation of random numbers and transferring of public value over the network. It results in a variable delay for KE which is why we have not included the latency values involving KE.

# 6. CONCLUSIONS

Securing embedded controller devices such as PLCs or RTUs against attacks by adding security mechanisms to these resource and memory constrained devices can mean sacrificing performance. Due to this trade-off, often security-related decisions are delayed from early design and development phases to the later deployment phase. This paper proposes an approach to make security related decisions. At the heart of this approach is a confidentiality-based security layer for function blocks called CL4FB, implemented using IEC 61499 FBs. This layer acts as a library for secure communications and supports a variety of security algorithms varying in performance overheads and levels of security provided. Through a case study of a smart-grid protection system, we show that even for communications constrained by strict hard real-time deadlines, some level of security can be incorporated.

Future directions for this research include the implementation of additional confidentiality, integrity and availability related mechanisms and algorithms to the library and their benchmarking. Finally, tool-support for automatically instantiating security blocks against secure FB links required for any IEC 61499 application is another promising research avenue that we are exploring.